\documentclass[twocolumn,prl,showpacs]{revtex4}
\usepackage[T1]{fontenc}
\usepackage[latin1]{inputenc}
\usepackage{amsmath}
\usepackage{graphicx}
\usepackage{amssymb}

\makeatletter
\usepackage{graphicx}
\usepackage{bm}

\makeatother
\begin{document}

\title{Canonical Bose gas simulations with stochastic gauges}

\author{P.D. Drummond, P. Deuar, K.V. Kheruntsyan}

\affiliation{ARC Centre of Excellence for Quantum-Atom Optics, Department of Physics, \\
 University of Queensland, Brisbane, Qld 4072, Australia}

\date{\today {}}

\begin{abstract}
A technique to simulate the grand canonical ensembles of interacting Bose gases is presented. Results are generated for many
temperatures by averaging over energy-weighted stochastic paths, each corresponding to a solution of coupled Gross-Pitaevskii
equations with phase-noise. The stochastic gauge method used relies on an off-diagonal coherent-state expansion, thus taking
into account all quantum correlations. As an example, the second order spatial correlation function and momentum distribution
for an interacting 1D Bose gas are calculated. 
\end{abstract}

\pacs{05.30Jp, 03.75.Hh, 03.75.Pp}

\maketitle
Calculating the observables corresponding to a grand canonical density matrix is a highly nontrivial problem in quantum many-body
theory, due to the large dimensionality of the corresponding Hilbert space. In this Letter, we show that for Bose gases,
this problem is soluble using a coherent-state stochastic gauge method, which generates equations similar to the widely used
Gross-Pitaevskii equation \cite{BECTheory}, with additional quantum noise terms. The issue is of much topical interest,
as it relates not just to recent experiment and theory on trapped dilute gas Bose-Einstein condensates (BEC) \cite{Nature-special-issue},
but also to fundamental questions like the fermion-boson duality problem \cite{Fermi-Bose} in quasi one-dimensional ($1D$)
Bose gas systems \cite{1D-Bose-gas-exp}.

A major advantage of the method presented is that the number of variables is linear in the number of spatial lattice points,
in any number of dimensions. This scaling suggests it is competitive in efficiency with Monte Carlo techniques, while being
able to in principle calculate arbitrary observables. Additionally, it is applicable to both imaginary time evolution equations
describing thermal equilibrium and to quantum dynamical calculations in real time. As a result, the same technique can be
used to calculate state preparation and subsequent time evolution. In this Letter we calculate equilibrium momentum distributions
and spatial density correlation functions for a finite-temperature $1D$ Bose gas, as well as verifying the technique by
comparisons with known exact results.

The exactly solvable $1D$ Bose gas model with repulsive delta function interactions \cite{Girardeau,LiebLiniger,Yang} has
been the subject of renewed theoretical interest recently, owing to experimental realization of the one-dimensional regime
in trapped alkali gases \cite{1D-Bose-gas-exp}. The higher-order correlation functions are responsible for the rates of
inelastic collisional processes, and are of particular importance for the studies of interference and coherence properties
of atom lasers operating in one-dimensional waveguide environments. Local higher-order correlations at zero and finite temperatures
have recently been calculated \cite{Gangardt-Shlyapnikov,KK-DG-PD-GS}, utilizing the previously known exact analytic solutions\cite{LiebLiniger,Yang}.

Here we employ the stochastic gauge method to obtain results for the \textit{non-local} second-order correlation function.
In particular, we investigate the signatures of the cross-over from the weak-coupling Gross-Pitaevskii (GP) regime to the
strong-coupling Tonks-Girardeau (TG) regime where the system behaves like a free Fermi gas. The results indicate that these
correlations are a stronger indicator of quantum effects than the recently observed momentum distributions \cite{Richard}.

The stochastic gauge method is a generalization of the positive $P$ -representation approach, and is described in detail
in Ref. \cite{gauge-Drummond}. The original $P$-function expansion in quantum optics using coherent states\cite{schrodinger:1926},
was due to Glauber and Sudarshan \cite{glauber:63}. This does not give a positive distribution for all density matrices
and was later generalized to \textit{non-diagonal} $P$-functions \cite{drummondgardiner:80,gardiner}, which always exist
and are positive. 

This positive $P$-representation has been successfully applied to quantum dynamical problems in quantum optics \cite{CDRS}
and many-body theory \cite{drummondcorney:99}. If there is no damping, the method has stability problems \cite{bterms}
which are overcome by introducing stochastic gauges \cite{gauge-Drummond}. Other known methods include the quantum Monte-Carlo
methods \cite{ceperley:99+Krauth:96}, and stochastic wave-function methods \cite{Carusotto,Plimak} which have been applied
to BECs. However, these cannot be used for grand canonical ensembles, nor for subsequent quantum dynamical calculations if
there are losses.

We start by considering the interacting Bose gas model with repulsive delta-function interaction between the particles. In
second quantization, the model Hamiltonian is \begin{align}
\hat{H} & =\int\left[\frac{\hbar^{2}}{2m}{\bm\nabla}\hat{\Psi}^{\dagger}({\bm x}){\bm\nabla}\hat{\Psi}({\bm x})\right.\nonumber \\
 & \left.+V({\bm x})\Psi^{\dagger}({\bm x})\hat{\Psi}({\bm x})+\frac{U}{2}\hat{\Psi}^{\dagger}({\bm x})^{2}\hat{\Psi}({\bm x})^{2}\right]d{\bm x.}\label{model}\end{align}
 Here $\hat{\Psi}({\bm x})$ is the field operator, $m$ is the mass, $U>0$ is the coupling constant for the inter-particle
interaction, and $V({\bm x})$ is an external potential.

Restricting ourselves to the $1D$ case, with the understanding that $U$ corresponds to the $1D$ coupling strength \cite{Olshanii,Gangardt-Shlyapnikov}
, let us reduce Eq. (\ref{model}) to a Bose-Hubbard lattice Hamiltonian \cite{BoseHubbard} which contains all the essential
features. For a lattice with $M$ sites and spacing $\Delta x$, defining lattice annihilation operators $\hat{\mathbf{a}}=\{\hat{a}_{j}\}=\{\sqrt{\Delta x}\hat{\Psi}(x_{j})\}$,
one obtains (summing over repeated indices): \begin{equation}
\hat{H}(\hat{\mathbf{a}},\hat{\mathbf{a}}^{\dagger})=\chi\left[:\hat{n}_{j}^{2}:+2\omega_{ij}\hat{a}_{i}^{\dagger}\hat{a}_{j}\right].\label{Hamiltonian}\end{equation}
 In this Hamiltonian, the $\omega_{ij}$ represent the dimensionless oscillator frequencies together with a linear coupling to
other sites, while $\chi=U/2\Delta x$ is the rescaled coupling. The boson commutators are $[\hat{a}_{i},\hat{a}_{j}^{\dagger}]=\delta_{ij}$,
and $\hat{n}_{j}=\hat{a}_{j}^{\dagger}\hat{a}_{j}$ is the boson number operator for the $j$-th site.

To extend the gauge $P$-representation equations\cite{gauge-Drummond} to imaginary time or temperature, consider that when
$[\hat{H},\hat{N}]=0$ , the un-normalized density matrix of the grand canonical ensemble is given by \begin{equation}
\hat{\rho}_{u}=e^{-(\hat{H}-\mu\hat{N})/k_{B}T}=e^{-\hat{K}\tau}.\label{GCE}\end{equation}
 Here $\mu(T)$ is the temperature-dependent chemical potential, $\hat{N}$ is the particle number operator, and we introduce
dimensionless `Kamiltonian' $\hat{K}(\mu,\hat{a}_{j},\hat{a}_{j}^{\dagger})=(\hat{H}-\mu\hat{N})/\chi$ and $\tau=\chi/(k_{B}T)$
corresponding to inverse temperature. In the Schr\"{o}dinger picture, Eq. (\ref{GCE}) leads to the {}``imaginary time''
master-like equation \begin{equation}
\frac{\partial\hat{\rho}_{u}}{\partial\tau}=\left[\mu_{e}\hat{n}_{j}-\omega_{ij}\hat{a}_{i}^{\dagger}\hat{a}_{j}-\frac{1}{2}:\hat{n}_{j}^{2}:\,,\hat{\rho}_{u}\right]_{+}.\label{MasterEquation}\end{equation}
 where $\mu_{e}=\partial\lbrack\tau\mu(\tau)]/\partial(2\chi\tau)$ and $[A,B]_{+}$ stands for anticommutator. The usefulness
of this equation relies on the fact that at initial {}``time'' $\tau=0$ (i.e. $T\rightarrow\infty$) the state is $\hat{\rho}_{u}(0)=\exp(-\lambda\hat{N})$,
where $\lambda=-\lim_{\tau\rightarrow0}[\tau\mu(\tau)]/\chi$, and the initial number of particles per site is $n_{0}=[\exp(\lambda)-1]^{-1}$.
This initial state can then be evolved in $\tau$ to obtain grand canonical ensembles at all other (lower) temperatures $T$.

We expand the density matrix with a distribution $G$ and an additional quantum amplitude $\Omega$: \begin{equation}
\hat{\rho}_{u}=\int G(\overrightarrow{\alpha})\widehat{\Lambda}(\overrightarrow{\alpha})d^{4M+2}\overrightarrow{\alpha},\end{equation}
 where $\widehat{\Lambda}=\Omega\,\left\Vert {\bm\alpha}\right\rangle \left\langle \bm\beta^{\ast}\right\Vert e^{-{\bm\alpha}\cdot{\bm\beta}}$,
${\bm\alpha}$ and ${\bm\beta}$ are complex $M$-dimensional vectors of Bargmann coherent state amplitudes with elements
at each lattice point, and the notation $\overrightarrow{\alpha}$ is shorthand for all the variables $(\Omega,{\bm\alpha},{\bm\beta})=(\alpha_{0},\alpha_{1},\dots,\alpha_{2M})$.
The initial gauge-distribution is: \begin{equation}
G_{0}(\overrightarrow{\alpha})\propto\exp\left[-\left|\bm{\alpha}\right|^{2}/n_{0}\right]\delta^{2M}(\bm{\alpha}-\bm\beta^{\ast})\delta^{2}(\Omega-1).\label{P0}\end{equation}

To determine the effects of the time evolution (\ref{MasterEquation}) we use standard differential identities \cite{gauge-Drummond}
to replace the annihilation and creation operators acting on the projector $\widehat{\Lambda}$ . Using these operator identities
the operator equation (\ref{MasterEquation}) can be transformed to: \begin{equation}
\frac{\partial\hat{\rho}_{u}}{\partial\tau}=\int G(\overrightarrow{\alpha})\,\,\mathcal{L}\widehat{\Lambda}\,\, d^{4M+2}\overrightarrow{\alpha}.\label{inteqn}\end{equation}
 We can define the resulting differential operator $\mathcal{L}$ in two parts as $\mathcal{L}=\mathcal{L}^{K}+\mathcal{L}^{g}$,
by using the operator identities to obtain $\mathcal{L}^{K}=-K(\tau)+\sum_{j=1}^{2M}\left[A_{j}^{K}\partial_{j}-\alpha_{j}^{2}\partial_{j}^{2}/2\right]\,$
-- and we will define the {}``gauge'' term $\mathcal{L}^{g}$ next. Here, $A_{i}^{K}=(\mu_{e}-n_{i})\alpha_{i}-\omega_{ij}\alpha_{j}$,
$\omega_{i+M,j+M}=\omega_{ji}, \omega_{i+M,j}=\omega_{i,j+M}=0$, and $\partial_{j}=\partial/\partial\alpha_{j}$. The effective $c$-number `Kamiltonian'
is $K(\tau)\equiv K(2\chi\mu_{e}(\tau)\,,\bm{\alpha},\bm{\beta})$, while the effective complex boson number is $n_{j}=n_{j}^{\prime}+in_{j}^{\prime\prime}=n_{j+M}\equiv\alpha_{j}\beta_{j}$.

Now let us introduce a suitable stochastic gauge operator $\mathcal{L}^{g}$. The $\partial_{0}$ identity \cite{gauge-Drummond}
$\Omega\partial_{0}\widehat{\Lambda}=\widehat{\Lambda}$ allows the use of arbitrary stochastic gauge functions. This is
similar to an electrodynamic gauge, in that it defines an infinite class of physically equivalent Fokker-Planck and hence
stochastic differential equations. To show how this is achieved, define $2M$ arbitrary complex gauge functions $\mathbf{g}=\{ g_{i}(\overrightarrow{\alpha},\overrightarrow{\alpha}^{\ast})\}$
to give a gauge-dependent differential operator $\mathcal{L}^{g}$ which satisfies $\mathcal{L}^{g}\widehat{\Lambda}=0$:
\begin{equation}
\mathcal{L}^{g}=\left[-K(\tau)+\frac{1}{2}\mathbf{g}^{2}\Omega\partial_{0}+i\sum_{j=1}^{2M}g_{j}\alpha_{j}\partial_{j}\right]\left(\Omega\partial_{0}-1\right).\end{equation}

With a suitable gauge choice that eliminates boundary terms, we can now transform the equation (\ref{inteqn}) into a positive-definite
\cite{gauge-Drummond,drummondgardiner:80} Fokker-Planck equation via partial integration, and then into Ito stochastic differential
equations using standard methods \cite{gardiner}: \begin{equation}
\frac{d\alpha_{j}}{d\tau}=A_{j}+i\alpha_{j}\zeta_{j}(t)+\delta_{j0}\Omega\sum_{i=1}^{2M}g_{i}\zeta_{i}(t)\,.\end{equation}
 The \emph{total} complex drift vector including the quantum amplitude $\Omega$ is now $\overrightarrow{A}=(-\Omega K(\tau),A_{1},\dots,A_{2M})$,
where for $j>0$, $A_{j}=A_{j}^{K}-ig_{j}\alpha_{j}$. Here $\zeta_{0}=0$, and $\zeta_{j}(t)$ are $2M$ independent Gaussian
noises such that $\langle\zeta_{i}(\tau)\zeta_{j}(\tau^{\prime})\rangle=\delta_{ij}\delta(\tau-\tau^{\prime})$.

With no gauge ($g_{i}=0$), one finds that for this system the equations are unstable for $n_{i}^{\prime}<0$ because they
contain terms of the form $\dot{\alpha_{i}}=-\alpha_{i}^{2}\beta_{i}$. These have been shown to lead to systematic errors
due to singular trajectories which cause boundary terms \cite{bterms}. We choose gauge functions of the form $g_{j}=i(n_{j}^{\prime}-|n_{j}|)$,
which remove the instabilities, and also stabilize the phase of $n_{j}$ so that $\beta_{j}\simeq\alpha_{j}^{\ast}$. This
then gives the stable equations \begin{align}
\frac{d\alpha_{i}}{d\tau} & =\left[\mu_{e}-|n_{i}|-in_{i}^{\prime\prime}\right]\alpha_{i}-\sum_{j=1}^{2M}\omega_{ij}\alpha_{j}+i\alpha_{i}\zeta_{i}(\tau),\nonumber \\
\frac{d\Omega}{d\tau} & =\left[-K(\tau)+\sum_{i=1}^{2M}g_{i}\zeta_{i}(\tau)\right]\Omega.\end{align}

There is an intuitive physical interpretation. Since $\beta=\alpha^{*}$ in the initial thermal state, each amplitude initially
obeys a Gross-Pitaevskii equation in imaginary time, with quantum phase-noise due to the interactions. This causes non-classical
statistics with $\bm{\alpha}\neq\bm\beta^{*}$ to emerge as the temperature is lowered. Along each path an additional ensemble
weight $\Omega$ is accumulated, which is proportional to the effective Gibbs factor $e^{-\int K(\tau)d\tau}$ for the path, together
with gauge-dependent terms. This gives a hybrid between a path integral and a purely stochastic type of simulation.

Observables, which are always expressible as sums of terms of the form $\hat{a}^{\dagger n}\hat{a}^{m}$, have expectation
values given by the stochastic averages \begin{equation}
\langle\hat{a}^{\dagger n}\hat{a}^{m}\rangle=\frac{\langle\beta^{n}\alpha^{m}\Omega+(\alpha^{n}\beta^{m}\Omega)^{\ast}\rangle_{\text{stoch}}}{\langle\Omega+\Omega^{\ast}\rangle_{\text{stoch}}}\end{equation}
 and can all be in principle calculated.

As a demonstration, we present the results of a uniform (untrapped) $1D$ Bose gas calculation in a regime which lies in
the transition region between the weakly- and strongly-interacting gas, where neither perturbation theories nor the TG Fermi
gas approximations work well.

The behavior of a uniform Bose gas described by the model (\ref{model}) in $1D$ depends on two parameters $T_{\text{rel}}=T/T_{d}$
and $\gamma=mU/\rho\hbar^{2}$. For an ideal gas with linear ($1D$) particle number density $\rho$, the transition from
Boltzmann to Bose statistics occurs around the quantum degeneracy temperature $T_{d}=2\pi\hbar^{2}\rho^{2}/mk_{B}$. This
corresponds to the temperature when the average spacing between particles $1/\rho$ equals the thermal de Broglie wavelength
$\Lambda_{\text{th}}=\sqrt{2\pi\hbar^{2}/mk_{B}T}$.

The second parameter of interest is the coupling strength $\gamma$. As $\gamma$ increases from zero, one moves from the
weakly-interacting GP regime, where the gas is well described by Bogoliubov theory, to a strongly interacting TG regime,
where it undergoes {}``fermionization'' \cite{Girardeau,LiebLiniger}. For example, at zero temperature the ground state
energy is described \cite{LiebLiniger} to within $\approx95\%$ accuracy by Bogoliubov theory for $\gamma\lesssim0.8$,
and by an ideal Fermi gas for $\gamma\gtrsim80$. Here we simulate for an intermediate regime with $\gamma=T_{\text{rel}}=10$,
which is more likely to be accessible experimentally.

A point to note is that the chemical potential of the diffusive reservoir in contact with the system $\mu(\tau)$, can be
chosen at will. Its form must fulfill two conditions:

(1) The desired values of $\gamma(T_{\text{rel}})$ are simulated. Since $\chi$ is a constant during the simulation, $\mu(\tau)$
must be tailored to give the desired densities $\rho(\tau)$, and hence $\gamma$.

(2) A finite lattice leads to a maximum momentum cutoff, so one must take care that this does not influence observables ---
i.e. that simulated momenta do not couple to momenta beyond the cutoff. Fortunately at high enough temperatures $T_{\text{rel}}\gg\gamma$,
kinetic processes dominate over interparticle interactions and the momentum modes decouple. Hence, it suffices to start the
simulation at $\tau_{0}$ in an ideal Bose gas state at a high-enough temperature so that $T_{\text{rel}}(\tau_{0})\gg\gamma(\tau_{0})$.
The initial state is then still given by a form like (\ref{P0}), but the initial occupations in momentum space $n_{0}(k)$
are proportional to the ideal gas Bose distribution $n_{0}(k)\propto\{\exp[(-\hbar^{2}k^{2}/2m+\mu)\tau_{0}/\chi]-1\}^{-1}$.
In the example considered here $T_{\text{rel}}(\tau_{0})/\gamma(\tau_{0})\simeq2.01\times10^{3}$ and this ratio is monotonically
decreasing with $\tau$.

To avoid wasteful calculation of empty modes in the simulation, the width of the momentum distribution $\propto\sqrt{T}\propto\rho\sqrt{T_{\text{rel}}}$
should be kept approximately constant throughout. Hence, we want $\rho(\mu(\tau))\propto1/\sqrt{T_{\text{rel}}(\tau)}$.
A convenient form of $\mu(\tau>\tau_{0})$ gives a smooth interpolation between an initial fugacity $z_{0}=\exp(\tau_{0}\mu(\tau_{0})/\chi)$,
and a final fugacity $z_{f}$. We choose the function: $2z(\tau)=z_{0}+z_{f}+(z_{f}-z_{0})\cos\left(\pi\lbrack\tau-\tau_{f}]/[\tau_{f}-\tau_{0}]\right)$
.

To obtain the desired $\gamma=T_{\text{rel}}=10$ at the target temperature $T_{f}=\chi/k_{B}\tau_{f}$ {[}criterion (1){]},
one must choose two of the three numerical parameters $\chi,\tau_{f}$, and $z_{f}$, while the third becomes a scaling parameter
which determines the choice of units. To also fulfill criterion (2) one must choose such values of $z_{0}$ and $\tau_{0}$
that $T_{\text{rel}}(\tau_{0},z_{0},\chi)\gg\gamma(\tau_{0},z_{0},\chi)$. We chose the values $z_{0}=z_{f}/1000$ and $\tau_{0}=\tau_{f}/4$.

\begin{figure}
\includegraphics[%
  width=8.5cm]{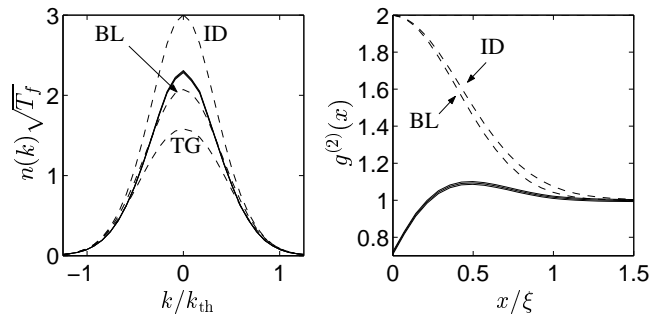}

\caption{$1D$ Bose gas in an intermediate regime: $T_{\text{rel}}=10$ and $\gamma=10$. Left panel shows the momentum distribution,
while the right panel is the second-order spatial correlation function. Here, the distance $x$ is in units of the healing
length $\xi$, which can be expressed as $\xi=\Lambda_{\text{th}}\sqrt{2T_{\text{rel}}/2\gamma}$. Solid lines are the results
of numerical simulations with 738200 stochastic trajectories. Dashed lines are the results for $\gamma=0$ ideal Bose gas
(ID), $\gamma\rightarrow\infty$ Tonks-Girardeau (TG) gas, and $T_{\text{rel}}\rightarrow\infty$ Boltzmann (BL) gas, shown
for comparison.}

\label{10ng}
\end{figure}

Figure~\ref{10ng} compares two important observables for the $\gamma=T_{\text{rel}}=10$ homogeneous system, with those of ideal
Bose ($\gamma=0$) and TG ($\gamma\rightarrow\infty$) gases under the same temperature and chemical potential. A comparison
is also made to the Boltzmann distribution which occurs at high temperatures $T_{\text{rel}}\rightarrow\infty$. The particle
number density in momentum space (classically of $k$-space width $k_{\text{th}}=\Lambda_{\text{th}}/\sqrt{2\pi}$) is intermediate
between the ideal gas and the TG regime. At high momenta, all the tails are seen to be Boltzmann-like, whereas for low momenta
Bose enhancement occurs above the Boltzmann level, but to a much smaller degree than in the ideal gas.

The second-order spatial correlation function \begin{equation}
g^{(2)}(x)=\frac{\langle\hat{\Psi}^{\dagger}(0)\hat{\Psi}^{\dagger}(x)\hat{\Psi}(x)\hat{\Psi}(0)\rangle}{\langle\hat{\Psi}^{\dagger}(0)\hat{\Psi}(0)\rangle\langle\hat{\Psi}^{\dagger}(x)\hat{\Psi}(x)\rangle}\end{equation}
 quantifies the spatial clumping of particles. A spatially uncorrelated field has $g^{(2)}(x>0)=1$. Some features that can
be seen from Fig. \ref{10ng} include: (\textit{i}) A peak arises at $x\simeq0.5\,\xi$, corresponding to the most likely
separation between particles. This effect is not present in an ideal gas. (\textit{ii}) Quite strong anti-bunching ($g^{(2)}(0)<1$)
of the bosons is seen: $g^{(2)}(0)=0.72\pm0.01$. (\textit{iii}) While the momentum distribution $\rho(k)$ is quite similar
to that of a Boltzmann gas, the behavior of the pair correlation function $g^{(2)}(x)$ is widely different.

\begin{figure}
\includegraphics[%
  width=8.5cm]{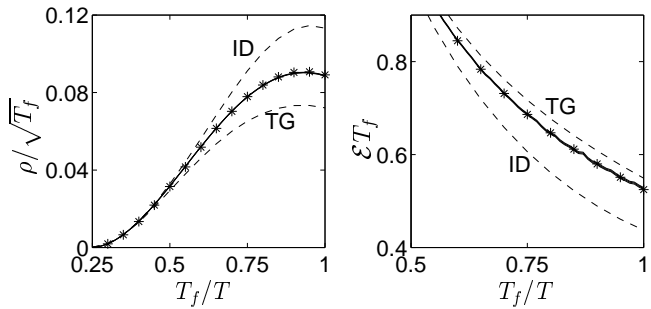}

\caption{Comparison of numerical calculation (solid lines) to exact results \cite{Yang} (asterisks), as well as to Ideal (ID) and
Tonks-Girardeau (TG) gases at the same temperature $T$ and chemical potential. Left panel: particle number density. Right
panel: energy per particle $\mathcal{E}$. }

\label{1pe}
\end{figure}

As an added test, we compared these calculations with exact results for boson density and energy derived by Yang and Yang
\cite{Yang}. This comparison is shown in Fig. \ref{1pe}, and one sees that the agreement is excellent. Similar good agreement
is found with the value of $g^{(2)}(0)$ calculated \cite{KK-DG-PD-GS} from the exact Yang-Yang theory.

In summary, we have applied the gauge $P$-representation technique to derive a set of stochastic equations whose averages
allow the calculation of all observables for the grand canonical ensemble of an interacting Bose gas modeled by Eq. (\ref{model}).
Results are obtained for a range of temperatures in one simulation. In this, an ensemble of trajectories representing a quantum
many-body system is cooled from an initially high temperature. The technique is readily applicable to more dimensions and
external potential by a straightforward generalization of the derivation outlined here. Further improvements are possible
through optimizing the basis (e.g. with generalized Gaussian states), and by using other gauges and algorithms such as Metropolis
sampling.

We have used this method to calculate the second-order spatial correlation function and momentum distribution for an interacting
$1D$ Bose gas, in a regime where both strong and weak coupling approximations are invalid, and no exact results are currently
available. Our results show that the pair correlation function is a sensitive indicator of the Bose-Fermi crossover, even
at temperatures above quantum degeneracy.

This research was supported by the Australian Research Council. Discussions with G. Collecutt are acknowledged.

\end{document}